\documentclass[superscriptaddress,aps,pra,twocolumn,showpacs,preprintnumbers,amsmath,amssymb,floatfix,10pt]{revtex4-1}

\usepackage{graphicx}
\usepackage{dcolumn}
\usepackage{bm}
\usepackage{color}

\def\d{\downarrow}
\def\u{\uparrow}
\def\U{\Uparrow}

\newcommand{\eq}[1]{Eq.~(\ref{#1})}
\newcommand{\fig}[1]{Fig.~\ref{#1}}
\newcommand{\bra}[1]{ \left\langle #1 \right |}
\newcommand{\ket}[1]{ \left |#1 \right\rangle}

\def \half {\tfrac{1}{2}}
\def \rpt {{\bar r_{\rm pt}}}
\def \rline {{\bar r_{\rm lin}}}

\def \half {\tfrac{1}{2}}

\def\ba{\begin{eqnarray}}
\def\ea{\end{eqnarray}}
\def\beq{\begin{equation}}
\def\eeq{\end{equation}}

\begin{document}

\title{Bilayer fractional quantum Hall states with ultracold dysprosium}

\author{N. Y. Yao}
\affiliation{Department of Physics, University of California, Berkeley, California 94720, USA}
\author{S. D. Bennett}
\affiliation{Department of Physics, Harvard University, Cambridge, Massachusetts 02138, USA}
\author{C. R. Laumann}
\affiliation{Department of Physics, University of Washington, Seattle, Washington 98195, USA}
\author{B. L. Lev} 
\affiliation{Department of Applied Physics, Stanford University, Stanford, California 94305, USA}
\affiliation{Department of Physics, Stanford University, Stanford, California 94305, USA}
\affiliation{E. L. Ginzton Laboratory, Stanford University, Stanford, California 94305, USA}
\author{A.V. Gorshkov}
\affiliation{Joint Quantum Institute, NIST/University of Maryland, College Park, Maryland 20742, USA}
\affiliation{Joint Center for Quantum Information and Computer Science, NIST/University of Maryland, College Park, Maryland 20742, USA}

\date{\today}

\begin{abstract}

We show how dipolar interactions between dysprosium atoms in an optical lattice can be used to obtain fractional quantum Hall states. In our approach, dysprosium atoms are trapped one atom per site in a deep optical lattice with negligible tunneling. Microwave and spatially dependent optical dressing fields are used to define an effective spin-1/2 or spin-1 degree of freedom in each atom. Thinking of spin-1/2 particles as hardcore bosons, dipole-dipole interactions give rise to boson hopping, topological flat bands with Chern number 1, and the $\nu = 1/2$ Laughlin state. Thinking of spin-1 particles as two-component hardcore bosons, dipole-dipole interactions again give rise to boson hopping, topological flat bands with Chern number 2, and the bilayer Halperin (2,2,1) state. By adjusting the optical fields, we find a phase diagram, in which the (2,2,1) state competes with superfluidity. Generalizations to solid-state magnetic dipoles are discussed.

\end{abstract}



\maketitle

\section{Introduction}

In addition to their fundamental importance, topological phases of matter may eventually enable the realization of fault-tolerant quantum computing \cite{nayak08b} and robust quantum state transfer \cite{yao13b}.  Thanks to unprecedented controllability and purity, synthetic atomic, molecular, and optical systems are rapidly gaining momentum in their ability to exhibit some of the most exotic topological phases \cite{galitski13, goldman14, lin09c, jotzu14, struck12, aidelsburger13, hafezi13, rechtsman13}. 
Dipolar systems, such as Rydberg atoms, polar molecules, and magnetic atoms, have recently attracted a particular degree of attention thanks to the strength of the interactions and the natural link between dipolar interactions and topology \cite{micheli06, yao12d,yao13c,kiffner13, maghrebi15, gorshkov13, peter14, syzranov14}. Indeed, thanks to the Einstein-de-Haas effect, dipolar interaction 
can convert internal angular momentum into angular momentum describing the rotation of the two interacting dipoles around each other. It is not surprising that the resulting rotation can be harnessed for generating effective gauge fields. However, nearly all dipolar topological literature to date focuses on interacting topological states arising from flat Chern bands with a Chern number equal to one. The exception is Ref.\ \cite{peter14}, which does consider  bands with higher Chern number, but which provides no path for making the bands sufficiently flat for realizing interacting topological phases. In this article, utilizing the magnetic atom dysprosium \cite{lu12}, we demonstrate for the first time how to create a topological flat band with Chern number 2 and to  utilize the resulting band-structure to realize the bilayer Halperin (2,2,1) fractional quantum Hall state \cite{halperin83}. By adjusting the magnetic field, we find a phase diagram, in which the (2,2,1) state competes with neighboring superfluids. More generally, our goal here is to demonstrate  the controllability and  potential of magnetic dipoles for simulating many-body phases; however, we emphasize that the constraints required by our approach are challenging for current generation experiments, and rather than an experimental blueprint, our approach should be viewed as a  general framework.

The remainder of the article is organized as follows. In section \ref{sec:c1}, we describe the effective Hamiltonian associated with a two-dimensional lattice of ultracold dysprosium atoms. We demonstrate that magnetic dipolar interactions mediate both long-range dynamics (hopping) and interactions. Microwave and optical radiation is used to break time-reversal symmetry and to control the specific nature of the atomic degrees of freedom. By tuning these dressing parameters, we realize the $\nu=1/2$ fractional Chern insulator and clarify its characteristics with a variety of numerical diagnostics. In section \ref{sec:NV}, we generalize our approach to solid-state magnetic dipoles. Specifically, we consider the example of Nitrogen-Vacancy defects in diamond and offer a route to sub-optical-wavelength resolution dressing via patterned dielectrics.
In section \ref{sec:c2}, we discuss how to realize a $C=2$ topological flat band with dysprosium, by considering an effective spin-1 atomic degree of freedom. Upon populating this band-structure with a finite density of interacting particles, we find a ground state that exhibits a Hall conductivity $\sigma_{xy} = 2/3$ consistent with the (2,2,1) Halperin state. In section \ref{sec:exp}, we elaborate on the experimental considerations and discuss the challenges using current technologies. Finally, in section \ref{sec:outlook}, we present a brief outlook. 

\begin{figure}[h!]
\begin{center}
\includegraphics[width = 0.99 \columnwidth]{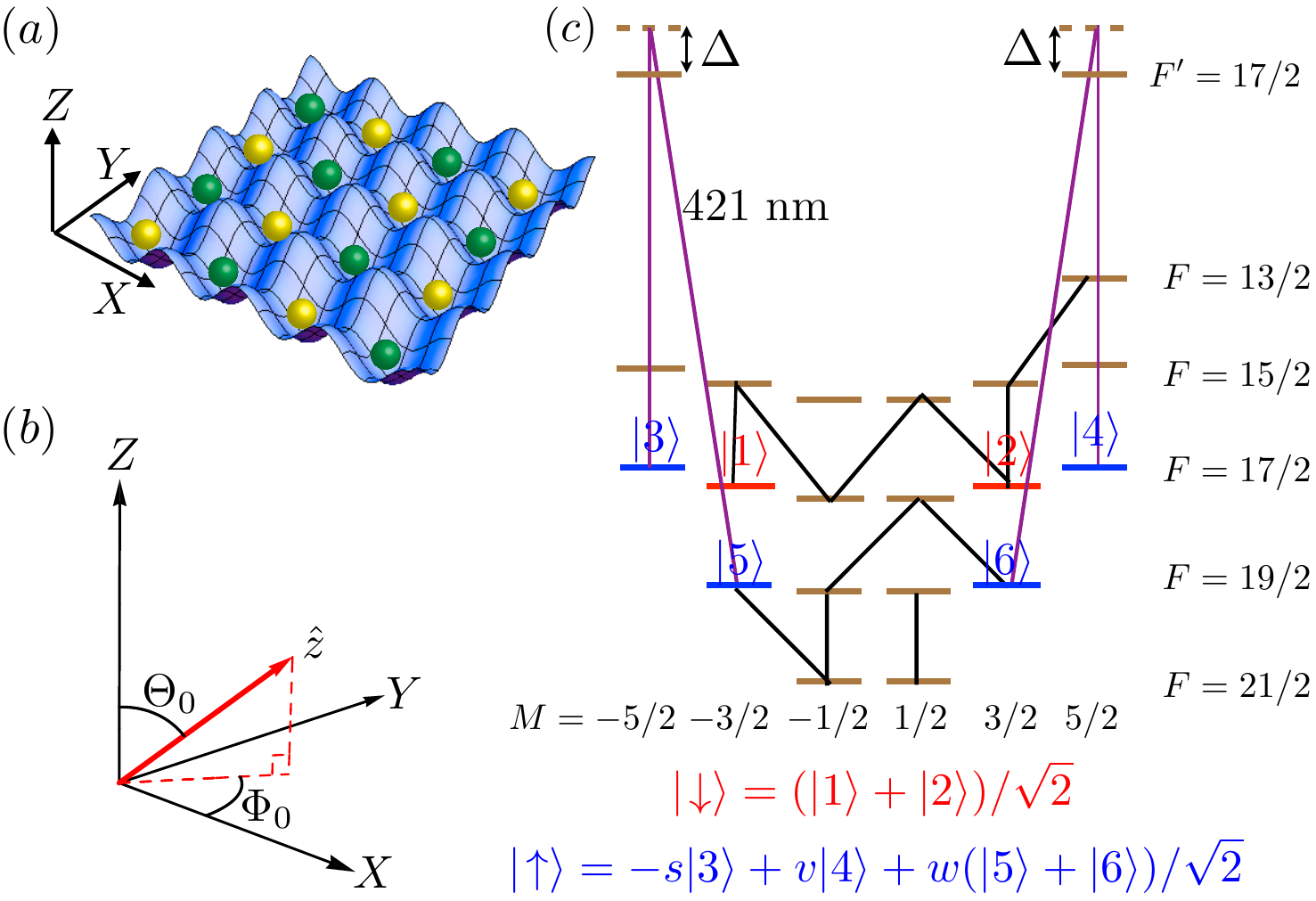} 
\caption{(a) Dysprosium atoms are loaded one atom per site into a square lattice in the $X$-$Y$ plane.  
(b) The quantization axis of the atoms $\hat z$ (red) is determined by the polarizations of applied fields. To obtain the $x$-$y$-$z$ coordinate system, one rotates the $X$-$Y$-$Z$ coordinate system by $\Phi_0$ around the $\hat Z$ axis and then by $\Theta_0$ around the $\hat y$ axis. (c) The relevant ${}^{161}$Dy level structure used for obtaining the flat band with Chern number $C=1$ and the corresponding $\nu = 1/2$ Laughlin state. The lower hyperfine levels (nuclear spin $I = 5/2$) are in the $J = 8$ ground electronic state.  Far-detuned light (not shown) provides a tensor AC Stark shift $\propto M^2$ with an $F$-dependent coefficient.  The 421 nm light (purple), which drives a transition to the $F' = 17/2$ hyperfine level of the $4f^{10}({}^5I_8)6s6p({}^1P^o_1)(8,1)^o_9$ state, and the microwaves (black) define the dark dressed states $\ket{\d}$ and $\ket{\u}$. The dressing optical field is different on the two sublattices [green and yellow in (a)] making the state $\ket{\u}$ sublattice-dependent.  \label{fig:scheme}}
\end{center}
\end{figure}

\section{$C=1$ flat band and the $\nu=1/2$ Laughlin state \label{sec:c1}}

To introduce the main features of dysprosium, as well as dipolar-mediated topological flat bands and fractional Chern insulators, we will first show how to obtain a flat band with Chern number $C = 1$ and use it to realize a $\nu = 1/2$ Laughlin state with dysprosium. As shown in Fig.\ \ref{fig:scheme}(a), we consider loading ${}^{161}$Dy  one atom per site in a square lattice in the $X$-$Y$ plane with nearest-neighbor spacing of $\lambda_\textrm{lat}/2 = 266$ nm created with off-resonant light of $\lambda_\textrm{lat} = 532$ nm wavelength. We further assume that the lattice is so deep that tunneling is negligible.  
As shown in Fig.\ \ref{fig:scheme}(b), the quantization axis $\hat z$, determined by the polarizations of applied fields, is pointing in a tunable direction different from $\hat Z$. An off-resonant linearly polarized optical field (not shown) provides a tensor shift $\propto M^2$ shown in Fig.\ \ref{fig:scheme}(c). Optical and microwave dressing fields are then used to construct dressed dark states $\ket{\d}$ and $\ket{\u}$, where the site-dependent coefficients $s$, $v$, and $w$ in the definition of $\ket{\u}$ are controlled by the amplitudes of the applied fields. We choose to work with $M$ near zero to maximize the transition dipole moments between the states involved and to simultaneously minimize the sensitivity of the energy levels to stray magnetic fields.

Interactions between the effective spin-$1/2$ particles are mediated by magnetic dipole-dipole interactions. Ignoring the small nuclear dipole moment, the interaction between two dysprosium atoms $i$ and $j$ separated by $(R,\theta,\phi)$ in the spherical coordinates associated with the $x$-$y$-$z$ coordinate system, in the units of $\mu_0 (g_J \mu_B)^2/(4 \pi R^3)$ (where  the Land\'e g-factor is $g_J = 1.24$ \cite{martin78}), is given by 
\ba
\hat H_{ij} 
&=& (1 - 3 \cos^2 \theta) \left[\hat J^z_i \hat J^z_j - \tfrac{1}{4} (\hat J^+_i \hat J^-_j + \hat J^-_i \hat J^+_j)\right] \nonumber \\
&& - \tfrac{3}{4} \sin^2 \theta \left[e^{- 2 i \phi} \hat J^+_i \hat J^+_j + h.c.\right]. \label{Jeq} 
\ea
Here $J=8$ is the total electronic angular momentum associated with the electronic ground state. We set $\hbar = 1$ throughout the paper. Assuming that the dark states $\ket{\u}$ and $\ket{\d}$ are separated from the bright states by an energy $\Omega_\textrm{dr}^2/\Delta$ (here $\Omega_\textrm{dr}$ and $\Delta$ being the Rabi frequency an detuning of the optical dressing fields; the microwave Rabi frequency is assumed to be $\sim \Omega_\textrm{dr}^2/\Delta$) that is larger than the dipole-dipole interaction strength, 
  we project the Hamiltonian onto the four-dimensional Hilbert space, where each of the two atoms is in state $\ket{\d} = \ket{0}$ or $\ket{\u} = \hat a^\dagger \ket{0}$. We have introduced the creation operator $\hat a^\dagger_j$ for a hardcore boson on site $j$. The Hamiltonian then takes the form
\ba
\hat H_\textrm{dd} =  \sum_{i \neq j} \left[t_{ij} \hat a^\dagger_i \hat a_j + V_{ij} \hat n_i \hat n_j\right],
\ea
where $\hat n_i = \hat a^\dagger_i \hat a_i$ and where the hopping amplitude $t_{ij}$ and the density-density interaction $V_{ij}$ (see Appendix \ref{sec:app1}) depend on the vector $(R, \theta, \phi)$ separating atoms $i$ and $j$, the direction $(\Theta_0,\Phi_0)$ of the quantization axis [see Fig.\ \ref{fig:scheme}(b)], the dressing coefficients $s_{i,j}$, $v_{i,j}$, $w_{i,j}$ [see Fig.\ \ref{fig:scheme}(c)], permanent dipole moment $\mu_4 = \langle 4|\hat J^z|4\rangle$ and transition dipole moments $\mu_{53}$, $\mu_{13}$, and $\mu_{26}$ (here $\mu_{\alpha \beta} = \langle \alpha|\hat J^{z,\pm}|\beta\rangle$, where the difference between the $M$ quantum numbers of $|\alpha\rangle$ and $|\beta\rangle$ determines the choice of $z$ or $\pm$). Notice that, in contrast to electric-dipole implementations, the bare $|F,M\rangle$ states in Fig.\ \ref{fig:scheme}(c) have substantial dipole moments even in the absence of applied fields. 

\begin{figure}[t]
\begin{center}
\includegraphics[width = 0.98\columnwidth]{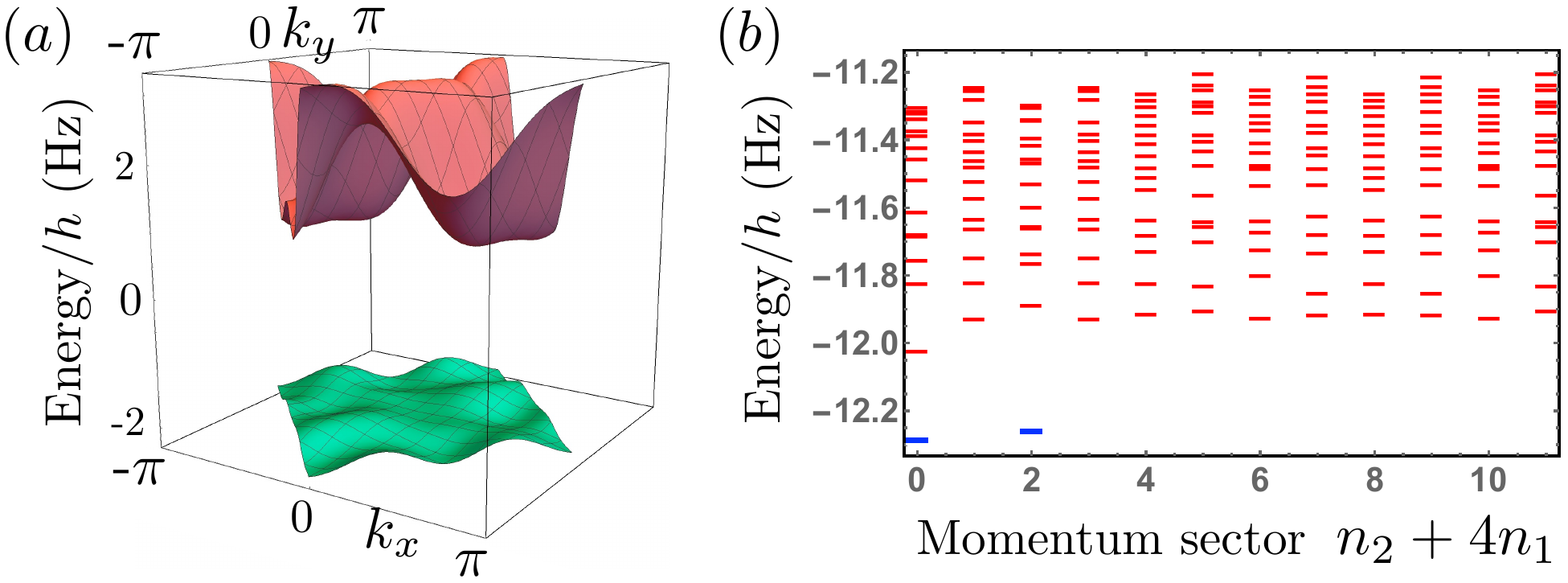}
\caption{(a) Flat topological band with Chern number $C = 1$. 
 The flatness of the band (band gap divided by band width) is $\approx 8$. 
  (b) The momentum-resolved eigenvalues for the $\nu = 1/2$ fractional Chern insulator with 6 hardcore bosons on a $6 \times 4$ torus. The momentum sector  $n_2 + 4 n_1$ 
corresponds to $(k_x,k_y) = (n_1/3,n_2/2-n_1/3) \pi$,  $n_1 = 0,1,2$, and $n_2 = 0,1,2,3$.
The spectrum features a gap separating two degenerate ground states at $(k_x,k_y) = (0,0)$ and $(0,\pi)$ (blue) from the other states by a gap. \label{fig:c1}}
\end{center}
\end{figure}

In Fig.\ \ref{fig:c1}, we show the $C=1$ topological flat band and the resulting $\nu = 1/2$ Laughlin state on a torus, exhibiting the expected gapped two-fold degenerate ground state. Each ground state was verified to have the manybody Chern number of 1/2 and the correct quasi-hole statistics obeying the generalized Pauli principle \cite{regnault11}. The specific values, used in Fig.\ \ref{fig:c1}, of site-dependent coefficients $s$, $v$, $w$ and of the direction of $(\Theta_0,\Phi_0)$ of the quantization axis relative to the $X$-$Y$-$Z$ plane are given in  Appendix \ref{sec:app1}.
 In particular, we would like to set arbitrary $s$, $v$, and $w$ on one sublattice [yellow in Fig.\ \ref{fig:scheme}(a)] and arbitrary $s$, $v$, and $w$ on the other sublattice [green in Fig.\ \ref{fig:scheme}(a)]. On top of that, we would like the value of $w$ to alternate every other row. The 421 nm optical fields shown in Fig.\ \ref{fig:scheme}(c) are key to generating this site dependence of $s$, $v$, and $w$, while the microwave fields provide spatially uniform couplings.
Specifically, first, we apply an optical field with a nonzero $\pi$-polarized component (with respect to the quantization axis $\hat z$) that is uniform across the square lattice by having the light's $k$-vector perpendicular to the plane of the atoms.  Furthermore, the beam's polarization can always be chosen such that its $\sigma^+$ and $\sigma^-$ components (with respect to the quantization axis $\hat z$) have equal intensity. These $\sigma^\pm$ components will thus only result in AC Stark shifts that keep the energies of $|F,M\rangle$ and $|F,-M\rangle$ equal and that can therefore be simply absorbed into the tensor shifts.
Second, since 421 nm is less than the wavelength of light $\lambda_\textrm{lat} = 532$ nm used to create the optical lattice,
we can take a beam propagating along $\hat X$ and rotate its $k$-vector slightly around $\hat Y$, such that, within the $XY$ plane of the atoms, it acquires periodicity of twice the lattice spacing along $\hat X$. We will have two such beams corresponding to clockwise and counterclockwise rotations around $\hat Y$. We then similarly take a beam propagating along $\hat Y$ and rotate its $k$-vector slightly around $\hat X$, such that, within the $XY$ plane of the atoms, it acquires periodicity of twice the lattice spacing along $\hat Y$. Again, we will have two such beams corresponding to clockwise and counterclockwise rotations around $\hat X$. The electric field of each of the resulting four beams is described by two complex numbers corresponding to the amplitudes of the two transverse polarization components. 
The resulting eight complex numbers can be generically tuned to get no $\pi$-polarization amplitude on any of the atoms and simultaneously arbitrary 
$\sigma^+$ and $\sigma^-$ amplitudes on the two sublattices subject to the desired sign alteration every other row.

\section{Implementation in NV centers \label{sec:NV}}

To depict the generality of our construction, we now consider magnetic dipoles associated with Nitrogen-Vacancy color centers in diamond. The NV center has received a tremendous amount of interest in recent years owing to the fact that its electronic spin can be  polarized, manipulated and optically detected under ambient conditions \cite{Childress2006b,Gaebel2006,Dutt2007,Neumann2008,Balasubramanian2009,Neumann2010,Neumann2010a}. Each NV center also harbors a localized nuclear spin, which exhibits extremely long coherence times \cite{Maurer2012}. Using a combination of these electronic and nuclear degrees of freedom, we will demonstrate the ability to realize topological flat bands. Our approach will be analogous to the previous section: namely, the use of microwave and optical fields to break time-reversal symmetry and to realize appropriately dressed eigenstates. 

\begin{figure}
\centering
\includegraphics[width=3.40in]{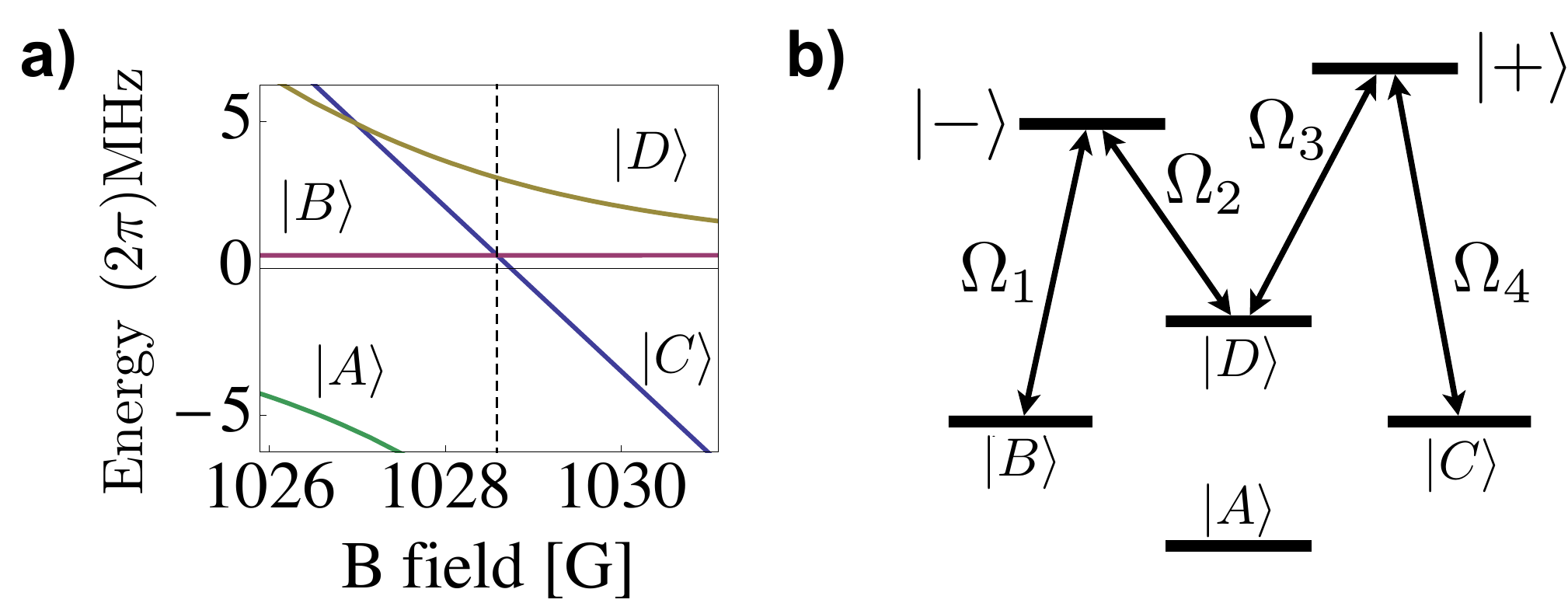}
\caption{ (a) Magnetic field required to tune the hyperfine coupled NV states to their desired resonances. (b) Optical dressing M-scheme  which enables sufficient control to realize topological flat bands.  \label{fig:nv1}} 
\end{figure}

The electronic ground state of each NV center 
is a spin-1 triplet described by the Hamiltonian,
\begin{equation}
\label{eq:Hnv}
	\hat H_{NV} = D_0 \hat S_z^2 + \mu_e B \hat S_z,
\end{equation}
where $D_0/2\pi = 2.87$ GHz is the zero
field splitting,
$\mu_e = - (2\pi) 2.8$ MHz/G is the
electron spin gyromagnetic ratio, and
$B$ is a magnetic field applied parallel
to the NV axis. This electronic spin is coupled via hyperfine interactions to the $I=1/2$ nuclear spin of the $^{15}$N impurity via
\begin{equation}
\label{eq:hyperfine}
	\hat H_{HF} =  A_{\parallel} \hat S_z \hat I_z + A_{\perp} (\hat S_x \hat I_x + \hat S_y \hat I_y),
\end{equation}
where $A_{\parallel}/2\pi \approx 3.0$ MHz and $A_{\perp}/2\pi \approx 3.7$ MHz \cite{Harrison2006,Gali2008,Felton2009}. 
We assume that a dc magnetic field tunes the energies of states $\ket{0,-\half}$ and $\ket{1,\half}$ to be nearly equal and simultaneously far-detunes the energies of states $\ket{-1,\pm \half}$, where
states are labeled by $\ket{S_z,I_z}$.
The $A_{\perp}$ term in \eq{eq:hyperfine} 
mixes the $\ket{0,\half}$ and $\ket{1,-\half}$ states, 
yielding the energy levels 
shown versus magnetic field in Fig.~\ref{fig:nv1}(a), where 
we have defined the
eigenstates 
$\ket{A} = \beta \ket{1,-\half}  -\alpha \ket{0,\half}$,
$\ket{B} = \ket{0,-\half}$,
$\ket{C} = \ket{1,\half}$,
and $\ket{D} = \alpha \ket{1,-\half} + \beta  \ket{0,\half}$. 
To allow for resonant hops of spin excitations, we work
at $B \approx 1028$ G where states
$\ket{B}$ and  $\ket{C}$ are nearly degenerate.

\vspace{5mm}

\noindent The effective states we use on each NV center are $\ket{0} = | A \rangle$ and $\ket{1} = s |B\rangle + v  |C\rangle + w  |D\rangle$. Analogous to the case of dysprosium, the coefficients $s,v,w$ are determined via an optical ``M'' dressing scheme [Fig.~\ref{fig:nv1}(b)] where the two
excited states are
$\ket{\pm} \propto \ket{E_x} \pm \ket{A_2}$, with
$\ket{E_x}, \ket{A_2}$ being two specific  electronic excited states of the NV \cite{Togan2010,Maze2011}. 
The state $\ket{1}$ is the so-called dark state of the M-scheme with $s=\Omega _2 \Omega _4/\tilde{\Omega} $, $v=\Omega _1 \Omega _3/\tilde{\Omega} $, $w=-\Omega _1\Omega _4/\tilde{\Omega} $. Note that lasers 1 and 3 must be linearly polarized,
while lasers 2 and 4 are circularly polarized. This elliptical polarization of light explicitly breaks time-reversal symmetry.

 The mixing angle $\tan (\theta_i) = |s_i/v_i|$ characterizes the strength of the effective dipole moment of  $\ket{1}$, thereby determining the magnitude of the interactions. In the limit  $\theta_i \rightarrow 0$, $s \rightarrow 0$; since the state $\ket{B}$ carries no  electronic spin dipole moment,  the dipolar interaction strength increases as $\theta_i \rightarrow 0$.  Topological flat-bands are found for a variety of parameter regimes, and fractional Chern insulating ground states are typically obtained for $\theta_i > 0.5$, where long-range interactions are relatively weak.
 
\subsection{Optical Dressing and Strain\label{sec:Efield}}

\begin{figure}
\centering
	\includegraphics[width=0.45\textwidth]{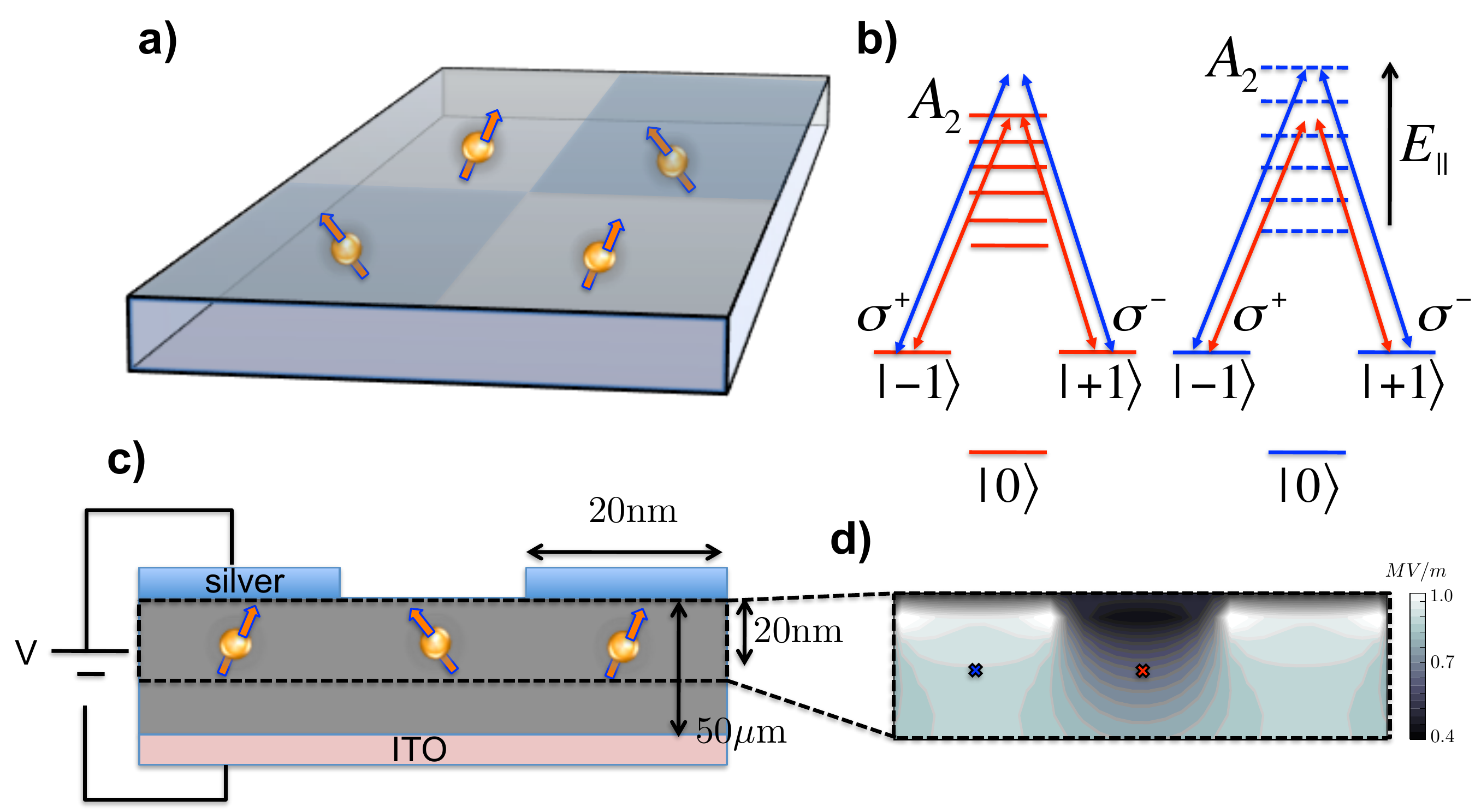}
\caption{
{ Control of solid-state spins.}
(a) Two dimensional array of implanted NV centers 
near diamond 110 surface, with NV lattice spacing
$a = 20$ nm.
A uniform applied electric field and
a  dielectric nanostructure patterned on the surface modulate
the displacement field at each NV center.
(b) Schematic optical dressing scheme. The left panel corresponds to the bare level structure and the right panel corresponds to the shifted level structure in the presence of gating. 
Spin states $\ket{\mp 1}$ in the electronic ground state
are coupled to the $\ket{A_2}$ excited state with 
$\sigma^{\pm}$ polarized light. We have left out the hyperfine structure for simplicity. 
Modulated displacement field brings the two frequencies (red and blue) of optical dressing in and out of resonance.
(c) Patterned dielectric on the surface of diamond (with an indium tin oxide back gate) which can yield a local dc field on the order of $10^3$V/cm as shown in (d). \color{black}
}\label{fig:idea}
\end{figure}

One challenge that arises in the context of implementing topological phases with solid-state magnetic dipoles is the ability to vary dressed states on   length scales smaller than an optical wavelength. Indeed, to generate to obtain  flat topological band structures, 
we require spatially inhomogeneous
optical dressing from site to site,
a nontrivial task for lattice spacings
$a \sim 20$ nm (required to achieve sufficiently strong magnetic dipole-dipole interactions) well below optical resolution.
In the case of NVs, one can, in principle, accomplish this task by
modulating an applied dc electric field
from site to site, using a 
patterned conducting nanostructure 
on the surface of the diamond as shown in 
\fig{fig:idea}.
A dc electric field applied
parallel to the NV axis shifts the electronic excited
state with respect to the ground state triplet due to its
strong electric dipole moment,
$d_{\parallel}^{ES} \sim (2 \pi) 1\ $MHz cm/V.
The ground state dipole moment is significantly weaker,
$d_{\parallel}^{GS} \sim (2 \pi) 0.35$Hz cm/V, and can be safely neglected \cite{Maze2011}. 
Thanks to the application of a dc voltage between the patterned surface conductor
and a back gate,
the electric field  on red and blue sites  tunes the optical transitions of the  NV centers 
in and out of resonance with red and blue optical driving
lasers, which are applied globally [Fig.~\ref{fig:idea}(b)].
For a diamond sample of thickness 10 $\mu$m, an applied voltage
of $V \sim 1$ V generates a dc field
of $E \sim 10^3$ V/cm [Fig.~\ref{fig:idea}(d), field difference between two x's], resulting in excited state shifts
of order $\sim$ GHz, which is significantly larger than the intrinsic line width $\gamma_e/2\pi \sim 10$ MHz. 

While this approach, in principle, allows for a site-dependent optical dressing, one needs to be extremely careful to consider the effects of local strain fields. These strain fields 
couples to NVs in the same way as an
effective electric field.
In diamond
grown by chemical vapor deposition (CVD)
and implanted with NV centers,
unwanted defects unavoidably lead to local variations in strain.
The two predominant sources
of strain in CVD diamond are point defects,
such as vacancy clusters and interstitials,
and line defects, such as stacking faults which
align along the growth direction.

To estimate the strain from point defects, we take
a defect concentration of 1 ppm \cite{Davies70, Collins92, Friel09},
corresponding to an average point defect separation of
$\rpt \sim 20$ nm.
For a defect strength of $A \sim 10^{-4}$ nm$^3$,
the resulting strain variations are of order
$\epsilon \sim A/\rpt^3 \sim 10^{-8}$, inducing
shifts of the NV excited state 
of $\sim 10$ MHz \cite{Eshelby57, Hardy68, Deshko13}.

Next we turn to
the expected dominant source of local strain variation, arising
from stacking faults aligned along the diamond growth direction.
We model the strain from stacking faults using elasticity theory
and assuming straight line defects.  
For a typical Burgers vector of magnitude $b \sim 2$ \AA,
the strain at  distance $\rline$ from the line defect is
given by $\epsilon \sim b\mu / \left[ 2\pi (1-\nu) \rline \right]$, 
where $\mu$ is the shear modulus
in diamond and $\nu$ is Poisson's ratio \cite{landau1986course}.
For an areal density of $10^4 / {\rm cm}^2$ \cite{TwitchenPC},
corresponding to an average distance of $\sim$ 100 $\mu$m
between line defects.
We estimate typical strain variations of 1-10 GHz
on length scales of 100 $\mu$m,
in agreement with recent experiments \cite{AlpPC}.
However, we expect much smaller strain variations 
on length scales smaller than $\rline$.
In particular, in a $250 \times 250$ nm region between line defects,
numerical simulations give average strain variations of $\sim 10$ MHz.

These estimates suggest that local strain variations in a small NV lattice might not destroy the spatially inhomogeneous optical dressing obtained via patterned electric fields; however, such an implementation is extremely challenging and our discussion of NV centers is meant mainly as a proof-of-principle analysis.

\section{$C = 2$ flat band and the Halperin (2,2,1) state \label{sec:c2}}

Having demonstrated the generality of our approach by extending it to NV centers in diamond, we now switch back to the dysprosium implementation. Having introduced in Sec.\ \ref{sec:c1} the potential of dysprosium for creating fractional quantum Hall states on the example of the $\nu = 1/2$ Laughlin state and the underlying $C=1$ flat band, we now move on to the construction of the topological flat band with Chern number $C=2$. At $\nu = 1/3$ filling fraction, this band will give rise to the Halperin (2,2,1) bilayer fractional quantum Hall state.

We will follow the idea of Ref.\ \cite{yang12} for generating the $C=2$ flat topological band. Suppose one has created a $C=1$ flat topological band on a square lattice with a two-site unit cell for a boson $\hat a$, like we did above for the hardcore bosons arising from the spin-1/2 model in dysprosium. Now let us introduce another species $\hat b$ of bosons that obeys exactly the same hopping Hamiltonian, except it is shifted relative to the hopping Hamiltonian for $\hat a$ by one lattice site in the $X$ direction. Diagonalizing the resulting Hamiltonian clearly gives four bands: two $C=1$ bands and two $C=-1$ bands. The idea is then to merge the two flat $C=1$ bands into a single flat $C=2$ band. To do this, on one sublattice, one defines the hardcore boson $\hat A = \hat a$ and the hardcore boson $\hat B = \hat b$, while on the other sublattice one does the opposite and defines $\hat A = \hat b$ and $\hat B = \hat a$. The resulting model recovers the full translational symmetry of the lattice and has therefore only two bands with $C = \pm 2$. We will realize the vacuum state and the two species of bosons on each site using a spin-1 particle. Therefore, our bosons will be hardcore both to themselves and to each other, interactions that will be sufficient for realizing the Halperin (2,2,1) bilayer state.

\begin{figure}[t]
\begin{center}
\includegraphics[width = 0.98\columnwidth]{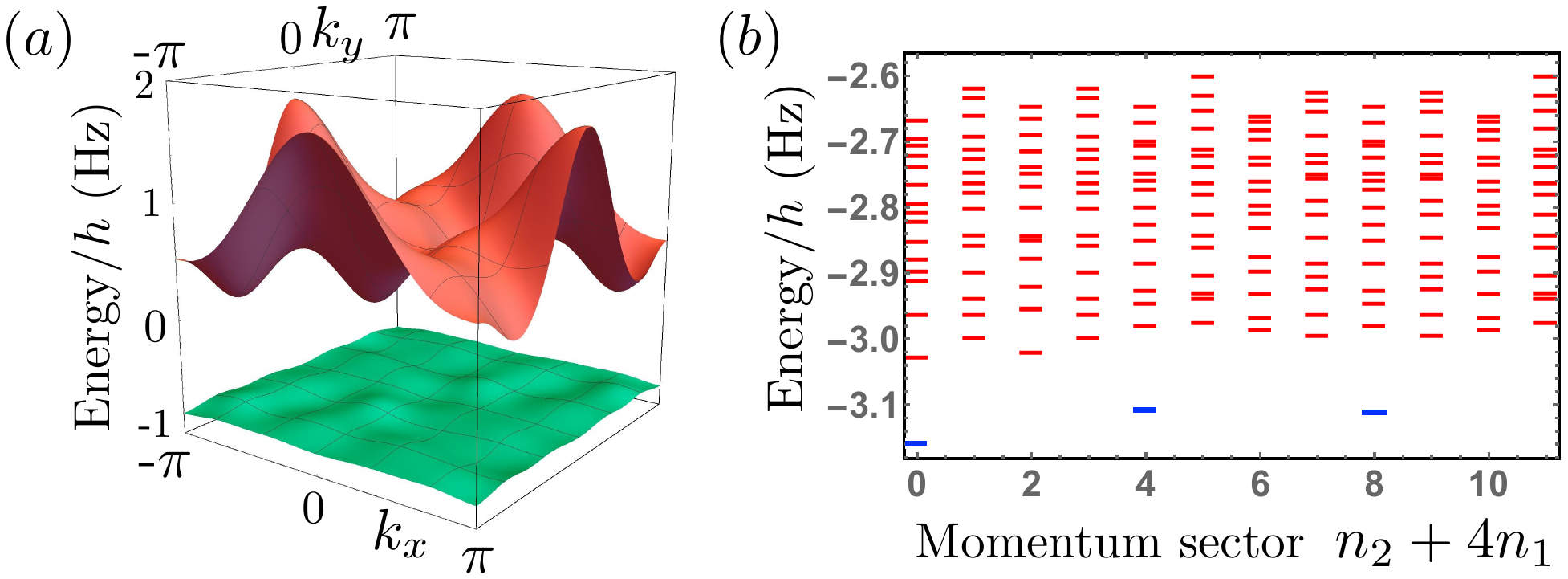}
\caption{
(a) Flat topological band with Chern number $C = 2$. 
The flatness of the band (band gap divided by band width) is $\approx 11$.  (b) The momentum-resolved eigenvalues for the $\nu = 1/3$    
fractional Chern insulator with 4 two-component hardcore bosons on a $3 \times 4$ torus. The momentum sector   
corresponds to $(k_x,k_y) = (2n_1/3,n_2/2)\pi$, 
$n_1 = 0,1,2$, 
and $n_2 = 0,1,2,3$. 
The spectrum features a gap separating the three degenerate ground states at $(k_x,k_y) = (0,0)$, ($2 \pi/3$,0), and ($4\pi/3$,0)  (blue) from the other states by a gap. \label{fig:c2}}
\end{center}
\end{figure}

We start with the same equation (\ref{Jeq}) as in the $C=1$ discussion, but now project onto dressed states $\ket{\d}$, $\ket{\u}$, and $\ket{\U}$, whose precise construction in terms of the ground hyperfine states of ${}^{161}$Dy is relegated to Appendix \ref{sec:app2}. We then define $\ket{0} = \ket{\d}$, $\hat a^\dagger \ket{0} = \ket{\u}$, and $\hat b^\dagger \ket{0} = \ket{\U}$. Choosing the dressed states 
in such a way that long-range density-density interactions vanish and such that $\hat a^\dagger$ and $\hat b^\dagger$ have the same hopping matrix elements $t'_{ij}$ (corresponding to a $C=1$ flat band) but shifted relative to each other by one unit in the $X$ direction, we obtain
\ba
\hat H_\textrm{dd} = \sum_{i \neq j} (t'_{ij} \hat a^\dagger_i \hat a_j + t'_{i-\hat X, j-\hat X} \hat b^\dagger_i \hat b_j).
\ea 
Following the above-described redefinition from $(\hat a, \hat b)$ to $(\hat A,\hat B)$, we arrive at a Hamiltonian describing a $C=2$ flat band. 
In Fig.\ \ref{fig:c2}(a), we show the resulting $C=2$ flat topological band. At $\nu = 1/3$ filling fraction, we will show that this topological flat band gives rise to the bilayer Halperin (2,2,1) state \cite{halperin83} shown in Fig.\ \ref{fig:c2}(b). 
To understand the nature of this state, it is helpful turn to the usual $K$-matrix description of bilayer quantum Hall systems, with wavefunction,
$\prod_{i<j} (z_{1i} - z_{1j})^l \prod_{i<j} (z_{2i} - z_{2j})^m  \prod_{i,j} (z_{1i} - z_{2j})^n e^{-\frac{1}{4} (\sum_i |z_{1i}|^2 + \sum_j |z_{2j}|^2 )}$
where $z_{pi}$ is the complex lattice coordinate of the $i$th hardcore boson in the $p$th layer. Here, $l$ and $m$ are even integers so that the wavefunction is consistent with Bose statistics, while $n$ can be any non-negative integer. The $K$-matrix is then defined as
$$K= \left[
   \begin{array}{ c c }
     l & n \\
     n & m
  \end{array} \right].
$$
In our case, from the structure factor, one finds that the state does not seem to preference a specific orbital type, suggesting a  charge vector $\vec{q} = (1,1)$. The transverse Hall conductance $\sigma_{xy}$ is then given by
\ba
\sigma_{xy} = qK^{-1}q.
\ea
From the numerics, we observe a gapped three-fold degenerate ground state on a torus,  each exhibiting a transverse Hall conductance, $\sigma_{xy} =  2/3$  \cite{yang12,wang12f}. Combined with the fact that ground state degeneracy is given by the determinant of $K$, this suggests that we are indeed observing the $(l,m,n) = (2,2,1)$ state.

\begin{figure}[t]
\begin{center}
\includegraphics[width = 0.7\columnwidth]{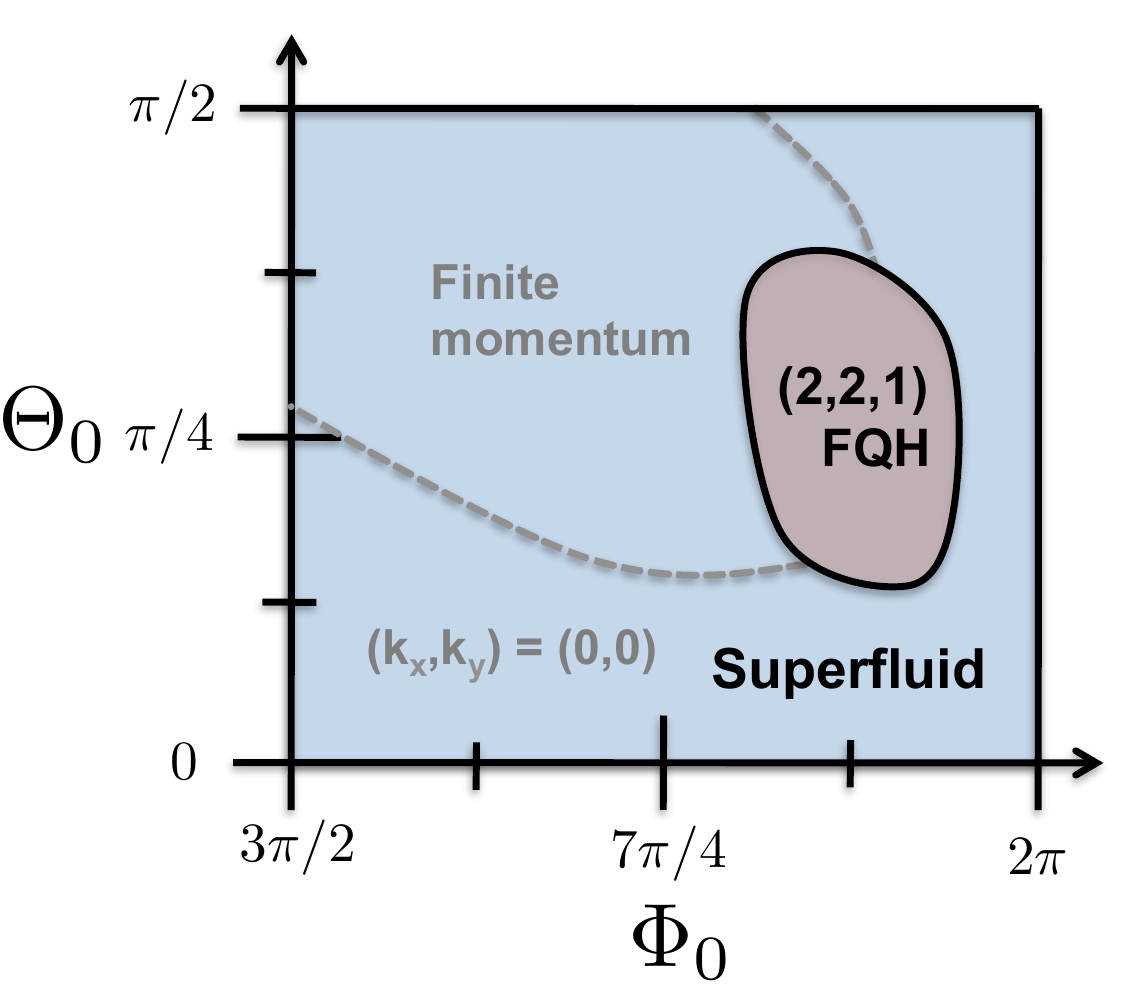}
\caption{The phase diagram obtained by starting with the Halperin (2,2,1) state shown in Fig.\ \ref{fig:c2} (red circle) and varying the direction ($\Theta_0$, $\Phi_0$) of the quantization axis.  \label{fig:c2diagram}}
\end{center}
\end{figure}

In Fig.\ \ref{fig:c2diagram}, we show the entire $C=2$ phase diagram [as a function of electric field tilt angle ($\Theta_0,\Phi_0$)], where the state in Fig.\ \ref{fig:c2}(b) is a single point (shown as a red circle) at $\Theta_0 = 0.68$ and $\Phi_0 = 5.83$. Similarly to Fig.\ \ref{fig:c2}, the phase diagram is obtained using exact diagonalization on a $3 \times 4$ torus, with two orbitals per site,  and a total of $4$ particles, corresponding to a density of $\nu=1/3$ particles per unit cell. 
The finite momentum superfluid has states which twist into the continuum but the ground state looks like it is at finite momentum instead of in the $(k_x,k_y) = (0,0)$ sector. 
The phase diagram is (a schematic smoothed version) based upon a back grid of 100 points. We sample at 10 equally spaced points along $\Theta_0$ and $\Phi_0$ and at each, we obtain the spectrum, the structure factor of the lowest energy state in all momentum sectors, the superfluid response from boundary condition twists, and the manybody Chern number $\sigma_{xy}$. In the region called (2,2,1), $\sigma_{xy}$ is numerically 2/3 within $< 1$\% error, which is a strong indicator that this is indeed the Halperin (2,2,1) state. 

\section{Experimental considerations \label{sec:exp}}

Our scheme relies on the following ladder of energy scales: $(V_\textrm{hf}/2\pi=1\textrm{ GHz}) \gg (V_\textrm{t}/2\pi \gg 300\textrm{ Hz}) \gg ((\Omega_\textrm{dr}^2/\Delta)/2 \pi = 30\textrm{ Hz})  
\gg (V_\textrm{dd}/2\pi= 3\textrm{ Hz}) \gg (E_\textrm{FQH}/2\pi = 500\textrm{ mHz}) \gg (\Gamma_\textrm{sc}/2 \pi = \Gamma'_\textrm{sc}/2 \pi = 100\textrm{ mHz})$.  The first inequality $V_\textrm{hf} \gg V_\textrm{t}$ ensures that the tensor AC Stark shift $V_\textrm{t}$ does not mix different $F$ levels separated by $V_\textrm{hf}$, which, in turn, ensures that the dipole moments of the involved states and transitions are given by their zero-field values, which simplifies the calculations. The second inequality $V_\textrm{t} \gg \Omega_\textrm{dr}^2/\Delta$ ensures that the optical fields (Rabi frequency $\Omega_\textrm{dr}$ and effective two-photon Rabi frequency $\Omega_\textrm{dr}^2/\Delta$) and microwave fields (Rabi frequency $\Omega_\textrm{dr}^2/\Delta$) are sufficiently weak to spectroscopically resolve the different transitions split by the tensor shift, allowing us to consider only the desired couplings. The inequality  $\Omega_\textrm{dr}^2/\Delta \gg V_\textrm{dd}$ ensures that dipole-dipole interactions $V_\textrm{dd}$, whose strength can be read out from the vertical energy scale  in Figs.\ \ref{fig:c1}(a) and \ref{fig:c2}(a), do not perturb the dark states $\ket{\u}$ and $\ket{\d}$ (and $\ket{\U}$ for the $C=2$ example) defined by the dressing fields and do not cause transitions from these dark states to the bright states.  We will discuss below that the small $3$ Hz interaction energy scale is not fundamental and can likely be significantly increased. The inequality $V_\textrm{dd} \gg E_\textrm{FQH}$, where $E_\textrm{FQH}$ is the fractional quantum Hall energy scale, which can be read out from the gap in Figs.\ \ref{fig:c1}(b) and \ref{fig:c2}(b), is not required but arises naturally since $E_\textrm{FQH}$ is determined by $V_\textrm{dd}$, so we included this inequality in the ladder of energy scales. Finally, the inequality  $E_\textrm{FQH} \gg \Gamma_\textrm{sc}, \Gamma'_\textrm{sc}$ ensures that the photon scattering rates $\Gamma_\textrm{sc}$ and $\Gamma_\textrm{sc}'$ due to the optical fields (used  for $\Omega_\textrm{dr}$ and for $V_\textrm{t}$, respectively) are much smaller than the energy scale $E_\textrm{FQH}$ of the Hamiltonian of interest.

Let us now discuss the scattering rate $\Gamma_\textrm{sc}'$ arising due to the optical fields creating $V_\textrm{t}$. In alkali atoms, the ratio of the tensor Stark shift $V_\textrm{t}$ to the scattering rate is small, and, for the case of Cs, maximized at only 8.2 (a fraction of the ratio between the hyperfine splitting and the linewidth), when the light is tuned between the $D_1$ hyperfine transitions \cite{chaudhury07}. On the other hand, the large orbital and spin angular momentum in the ground state of dysprosium gives rise to qualitatively stronger tensor light shifts and, hence, to much more favorable ratios of $V_\textrm{t}$ to the scattering rate \cite{cui13}. In particular, the tensor shift is on the order of $V_\textrm{t} = \Omega'^2/\Delta'$, while the decay rate is on the order of $\Gamma'_\textrm{sc} = \Gamma' (\Omega'/\Delta')^2$.  Here $\Omega'$, $\Delta'$, and $\Gamma'$ are the corresponding Rabi frequency, detuning, and spontaneous emission rate. Assuming we use 741 nm light, 
$\Gamma'/2\pi =  1.8$ kHz \cite{lu11}. To get $\Gamma'_\textrm{sc}/2\pi 
= 100\textrm{ mHz}$ and 
$V_\textrm{t}/2\pi = 300\textrm{ Hz}$, we need $\Delta'/2 \pi = 5\textrm{ MHz}$ and $\Omega'/2\pi = 40\textrm{ kHz}$, which is easily achievable. 

We now turn to the scattering rate $\Gamma_\textrm{sc}$ associated with the fact that dipole-dipole interactions make the dark state imperfect on the 421 nm transition. This scattering rate is given by $\Gamma_\textrm{sc} = \Gamma (V_\textrm{dd}/\Omega_\textrm{dr})^2$, where $\Gamma/2 \pi = 32\textrm{ MHz}$ is the linewidth of the 421 nm transition \cite{lu11}.  
To get $\Gamma_\textrm{sc}/2\pi = 100$ mHz and $\Omega_\textrm{dr}^2/\Delta = (2 \pi) 30$ Hz, we need $\Delta/2 \pi = 100$ MHz and $\Omega_\textrm{dr}/2 \pi = 50$ kHz, which is easily achievable. Assuming we are tuning to the blue of the $F' = 17/2$ hyperfine level of the excited state [see Fig.\ \ref{fig:scheme}(c)], the resulting detuning from both the $F' = 15/2$ state and the $F' = 19/2$ state is $\delta \approx (2 \pi) 1$ GHz, resulting in scattering from these levels at rates $\sim \Gamma (\Omega_\textrm{dr}/\delta)^2 < (2\pi) 100$ mHz.

An important advantage of our proposal is that we are not relying on collisions between the atoms, which allows us to avoid dipolar relaxation \cite{burdick15}. On the other hand, the main limitation of our proposal is the small energy scale $E_\textrm{FQH}/2\pi = 100$ mHz and the resulting stringent requirements on the linewidths of the lasers and coherence times of the atoms. An additional requirement is to make sure stray magnetic fields are weak enough that the Zeeman shifts are small on the scale of dipole-dipole interaction strength $V_\textrm{dd}/2\pi = 5$ Hz to ensure that the crucial $J^+_i J^+_j$ transitions in Eq.\ (\ref{Jeq}) stay resonant. 
In particular, this requires magnetic fields smaller than a micro Gauss, which is challenging. At the same time, our goal in the present manuscript is to demonstrate the controllability and the potential of magnetic dipoles in general, and magnetic atoms in particular, for obtaining fractional quantum Hall states in general and those in $C=2$ bands, in particular. We leave it to future work to optimize and modify the presented schemes with the goal of increasing the interaction energies by working at smaller lattice spacing  \cite{yi08,gullans12,romero-isart13,gonzalez-tudela14} and by choosing more optimal level configurations. Indeed, the maximum achievable interaction for two dysprosium atoms $\lambda_\textrm{lat}/2$ apart is  $\approx \mu_0 (10 \mu_B)^2/[4 \pi (\lambda_\textrm{lat}/2)^3] = (2 \pi) 70$ Hz and occurs when they are both in a $|J_z| = 8$ state. Therefore, optimization of our scheme can likely significantly raise $V_\textrm{dd}$ from its current value of $(2\pi) 3$ Hz.

To prepare the fractional Chern insulator ground state, we envision first tuning $\Theta_0$ and $\Phi_0$ to the part of the phase diagram where the ground state is a superfluid. If the superfluid phase exhibits weak quantum correlations, then it could, in principle, be approximated by a disentangled state, in which each atom is in a well-defined spin state. 
This can be prepared by adiabatically turning on the fields responsible for the tensor shift and the dressing on a time-scale faster than the inverse of $V_\textrm{dd}$.  
Next, we imagine adiabatically ramping from our prepared low-energy-density superfluid to the fractional Chern insulator. This can only be done if the phase transition between the two states is continuous \cite{barkeshli14b}, a question which is unknown for the (2,2,1) state.

The detection of the fractional Chern insulator can be carried out by effectively realizing Bragg spectroscopy. Specifically, the momentum- and energy-resolved spectral function can be measured by attempting to drive the $\ket{\d}-\ket{\u}$ transition with a different spatially dependent Rabi frequencies and different detunings. One can then use the spectral function to identify the bulk gap and the gapless chiral edges \cite{kjall12,goldman12b}.

\section{Outlook \label{sec:outlook}}

Our proposal generalizes naturally to other magnetic atoms, such as erbium \cite{aikawa12} and, to a lesser degree due to a smaller dipole moment, chromium \cite{paz13}. It can also be extended to engineer bilayer fractional Chern insulators in electric dipoles, such as Rydberg atoms and polar molecules.

This proposal also opens up avenues for engineering other exotic topological states. In particular, by replacing hopping terms (of the form $\hat a^\dagger_i \hat a_j + \textrm{h.c.}$)  with pairing terms (of the form $\hat a^\dagger_i \hat a^\dagger_j + \textrm{h.c.}$) along the rungs of a finite ladder, one expects to get parafermionic zero modes at the ends of the ladder, by analogy with proposals in solid-state heterostructures  \cite{clarke13,lindner12,cheng12,vaezi13} and cold atoms \cite{maghrebi15b}. The local replacement of pairing with hopping can be engineered by dressing the dipoles along the legs of the finite ladder in a way that is different from the rest of the system. An introduction of an array of such ladders into the (2,2,1) state may then lead, by analogy with Ref.\ \cite{mong14}, to a gapped ground state supporting Fibonacci anyons, which are universal for topological quantum computing via braiding alone \cite{nayak08b}. In fact, by extending the conjecture of Refs.\ \cite{mong14,vaezi14,vaezi14}, partially supported by numerics \cite{stoudenmire15,liu15} (although see Refs.\ \cite{geraedts15,peterson15}), one might expect that even the  introduction of uniform pairing on top of the (2,2,1) state may give rise to the Fibonacci phase. Instead of introducing pairing, one might also be able to obtain the Fibonacci state by introducing uniform  tunneling between two fractional Chern insulator layers \cite{vaezi14,liu15}, something we can also engineer in the dipolar approach. The use of uniform pairing or tunneling is an exciting prospect as the corresponding dressing would be significantly simpler relative to the approach involving an array of ladders \cite{mong14} and would therefore constitute a particularly promising approach towards universal topological quantum computing in ultracold atomic and molecular systems.

\section{acknowledgements}

We thank M.\ Lukin, 
N.\ Lindner,  
M.\ Maghrebi, N.\ Grabon, D.\ Clarke, and J.\ Alicea for discussions. This work was supported by the AFOSR, the Miller Institute for Basic Research in Science, NSF PIF, ARO, ARL, the NSF, the NSF PFC at the JQI,  AFOSR MURI, and the CUA.

\appendix

\renewcommand{\theequation}{A\arabic{equation}}
\renewcommand{\thefigure}{A\arabic{figure}}
\setcounter{equation}{0}
\setcounter{figure}{0}

\section{Details behind the $C=1$ topological flat band and the $\nu = 1/2$ Laughlin state \label{sec:app1}}

In this Appendix, we present the details behind the $C=1$ topological flat band and the $\nu = 1/2$ Laughlin state discussed in Sec.\ \ref{sec:c1}.

Assuming the tensor Stark shift $V_\textrm{t}$ is greater than the typical strength $V_\textrm{dd}$ of dipole-dipole interactions, we can project the Hamiltonian in Eq.\ (\ref{Jeq}) on the 36-dimensional Hilbert space, where each atom is in one of the states $\ket{1}$ through $\ket{6}$ defined in Fig.\ \ref{fig:scheme}(c), to obtain
\begin{widetext}
\begin{eqnarray}
 \hat H_{ij} 
&=& (1 - 3 \cos^2 \theta) \Big[\mu_4^2 (\hat n_4 - \hat n_3)(\hat n_4 - \hat n_3) 
+ \mu_{26}^2 (\ket{15}\bra{51}  +\ket{26}\bra{62} +\ket{16}\bra{52} + \ket{25}\bra{61} + h.c.) \nonumber \\
&&   \quad \quad \quad \quad \quad \quad - \tfrac{1}{4} \mu_{13}^2 ( \ket{13}\bra{31} + \ket{24}\bra{42} + h.c.)  - \tfrac{1}{4} \mu_{53}^2 ( \ket{53}\bra{35} + \ket{64}\bra{46} + h.c.)  \Big] \nonumber \\
&& - \tfrac{3}{4} \sin^2 \theta \left[e^{-2 i \phi}  \{\mu_{13}^2 (\ket{32}\bra{14} + \ket{23}\bra{41}) -\mu_{53}^2 (\ket{36}\bra{54} + \ket{63}\bra{45}) \}+ h.c.\right],
\end{eqnarray}
\end{widetext}
where $\mu_4 = \langle 4|\hat J^z|4\rangle = \frac{720}{323}$, $\mu_{26} =  \langle 2|\hat J^z|6\rangle = \frac{2 \sqrt{154}}{19}$, $\mu_{13} = \langle 1|\hat J^+|3\rangle = \frac{288 \sqrt{77}}{323}$ 
and $\mu_{53} = \langle 5|\hat J^+|3\rangle= - \frac{14 \sqrt{2}}{19}$. 

Assuming the separation $\Omega_\textrm{dr}^2/\Delta$ between the dark states $\ket{\u}$, $\ket{\d}$ and the bright states is larger than the interaction strength $V_\textrm{dd}$, we can further project the resulting Hamiltonian on the 4-dimensional Hilbert space spanned by $\ket{\d} = \ket{0}$ and  $\ket{\u} = \hat a^\dagger \ket{0}$  on each atom:
\begin{widetext}
\ba
\hat H_{ij} 
&=& (1 - 3 \cos^2 \theta) \Big[\hat n_i \hat n_j \left\{ \mu_4^2 (|v_i|^2 - |s_i|^2)(|v_j|^2 - |s_j|^2) - \tfrac{1}{8} \mu_{53}^2 (w_i^* w_j (s_i s_j^* + v_i v_j^*) + h.c.)\right\} \nonumber \\
&&  \quad \quad \quad \quad \quad \quad + \left\{\hat a^\dagger_i \hat a_j  ( \mu_{26}^2 w_i^* w_j  - \tfrac{1}{8} \mu_{13}^2 (s_i^* s_j + v_i^* v_j)) + h.c.\right\}\Big] \nonumber \\
&& - \tfrac{3}{4} \sin^2 \theta \left[e^{-2 i \phi}  \left\{ \tfrac{1}{2} \mu_{53}^2 \hat n_i \hat n_j (s_i^* w_j^* v_j w_i +s_j^* w_i^* v_i w_j)  - \tfrac{1}{2} \mu_{13}^2 (\hat a^\dagger_i \hat a_j v_j s_i^* + \hat a_i \hat a^\dagger_j v_i s_j^*)\right\}+ h.c.\right],
\ea
\end{widetext}
where $\hat n_j = \hat a^\dagger_j \hat a_j$.

In Fig.\ \ref{fig:c1}, we use $\Theta_0 = 2.35$ and $\Phi_0 = 4.13$ as the angles determining the direction of the DC electric field. The remaining parameters on the two sublattices (which we denote by  $1$ and $2$ in the subscript) are parametrized as $s_{1/2} = \sin(\alpha_{1/2}) \sin(\theta_{1/2})$, $v_{1/2}= \sin(\alpha_{1/2}) \cos(\theta_{1/2}) e^{i \phi_{1/2}}$, $w_{1/2} = \cos(\alpha_{1/2}) e^{i \gamma_{1/2}}$ with $\{\theta_1, \theta_2, \phi_1, \phi_2, \alpha_1, \alpha_2, \gamma_1, \gamma_2\} =  \{2.28, 1.59, -0.05, 1.51, 
0.46, 0.11, 1.60, 1.23\}$. As explained in the main text, we make a further modification by changing the sign of $w_{1}$ and $w_{2}$ every other row. While this modification does not increase the size of the unit cell, it is an important ingredient allowing us to obtain a sufficiently flat topological band.

In passing, we note that we can get the same bandstructure as for polar molecules \cite{yao13c} in a small electric field by adding two additional levels to the linear superposition composing $\ket{\u}$.

\section{Details behind the $C=2$ topological flat band, the Halperin (2,2,1) state, and the corresponding phase diagram \label{sec:app2}}

\begin{figure*}[t!]
\begin{center}
\includegraphics[width = 1.96 \columnwidth]{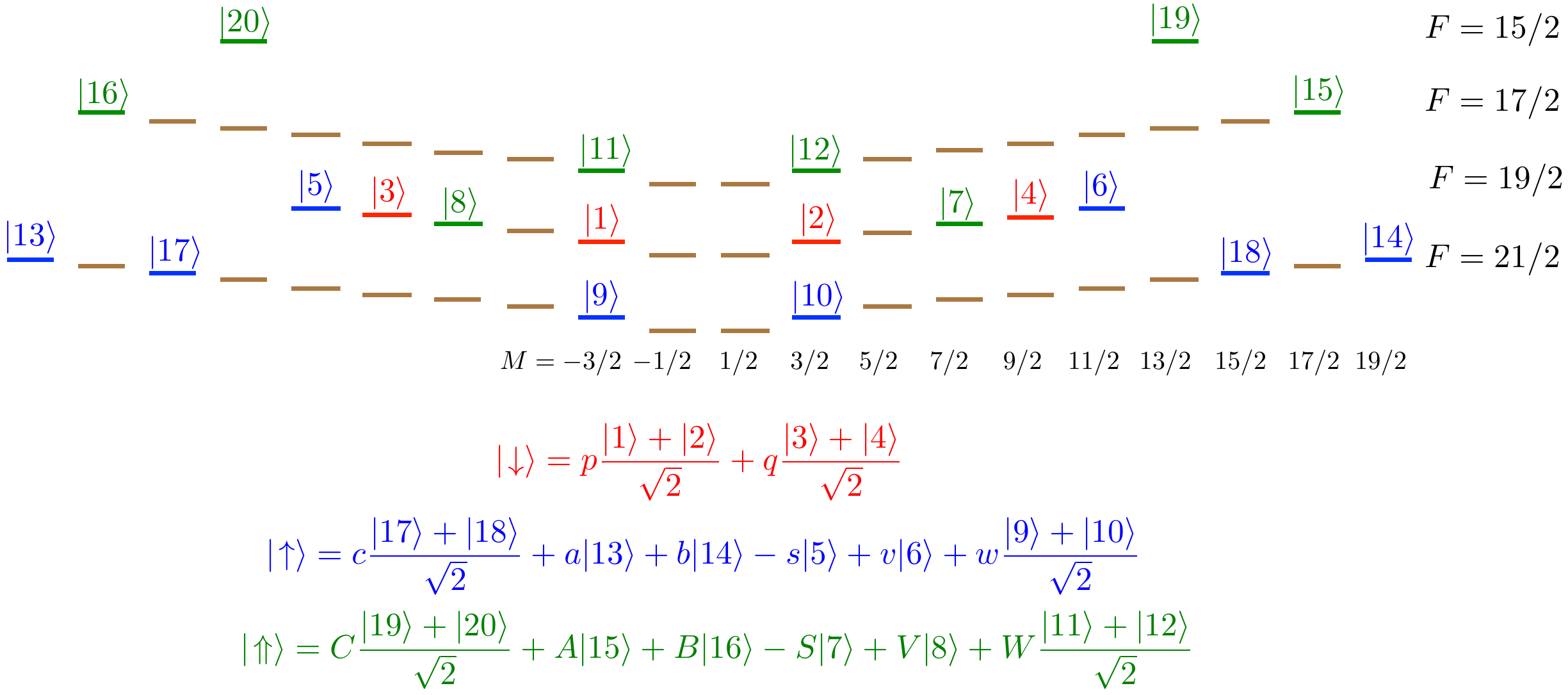}
\caption{The level structure of${}^{161}$Dy relevant for implementing the $C=2$ flat band and the resulting Halperin (2,2,1) state. \label{fig:scheme2}}
\end{center}
\end{figure*}

In this Appendix, we present the details behind the $C=2$ topological flat band and the Halperin (2,2,1) state discussed in Sec.\ \ref{sec:c2}.

Instead of using the simplest possible level structure for obtaining a flat $C=2$ topological band with ${}^{161}$Dy, we instead choose to demonstrate the full power and the full tunability of dipolar interactions between dressed states 
by realizing a Hamiltonian describing two-component hardcore bosons, each of which obeys the same Hamiltonian as hardcore bosons in $C=1$ bands realized with polar molecules \cite{yao13c}.
This derivation allows us to establish a certain degree of equivalence between implementations of topological flat bands and fractional quantum Hall states with different dipoles, both electric and magnetic. We will therefore consider the level structure and the dressed states shown in Fig.\ \ref{fig:scheme2}, where we assume $p$ and $q$ are real, positive, and site-independent.

Transition matrix elements between $\ket{\Uparrow}$ and $\ket{\u}$ vanish because of the $F \rightarrow F \pm 0,1$ selection rule ensuring that absence of terms exchanging the two types of excitations $\ket{\U}$ and $\ket{\u}$. Starting with Eq.\ (\ref{Jeq}), and keeping only those terms that will contribute to the interaction involving states $\ket{\d}$, $\ket{\u}$, and $\ket{\U}$, we obtain
\begin{widetext}
\ba
 \hat H_{ij} 
&=& (1 - 3 \cos^2 \theta) \Big[(\mu_6 (\hat n_6 - \hat n_5)  + \mu_7 (\hat n_7 - \hat n_8)+ \mu_{14} (\hat n_{14} - \hat n_{13})+ \mu_{16} (\hat n_{16} - \hat n_{15})) \times \nonumber \\
&&\quad \quad \quad \quad \quad \quad  \times (\mu_6 (\hat n_6 - \hat n_5)  + \mu_7 (\hat n_7 - \hat n_8)+ \mu_{14} (\hat n_{14} - \hat n_{13})+ \mu_{15} (\hat n_{15} - \hat n_{16})) \nonumber \\
&&\quad \quad \quad \quad \quad \quad  + \mu_{19}^2 (\ket{19}\bra{91}  +\ket{2,10}\bra{10,2} +\ket{1,10}\bra{92} + \ket{29}\bra{10,1} + h.c.) \nonumber \\
&&\quad \quad \quad \quad \quad \quad  + \mu_{1,11}^2 (\ket{1,11}\bra{11,1}  +\ket{2,12}\bra{12,2} +\ket{1,12}\bra{11,2} + \ket{2,11}\bra{12,1} + h.c.) \nonumber \\
&&   \quad \quad \quad \quad \quad \quad - \tfrac{1}{4} \mu_{35}^2 ( \ket{35}\bra{53} + \ket{46}\bra{64} + h.c.)  - \tfrac{1}{4} \mu_{83}^2 ( \ket{38}\bra{83} + \ket{47}\bra{74} + h.c.)  \Big] \nonumber \\
&& - \tfrac{3}{4} \sin^2 \theta \left[e^{-2 i \phi}  \{\mu_{35}^2 (\ket{36}\bra{54} + \ket{63}\bra{45}) +\mu_{83}^2 (\ket{84}\bra{37} + \ket{48}\bra{73}) \}+ h.c.\right]\\
 &=&(1 \!-\! 3 \cos^2 \theta) \Big[(\hat n_i \mu_{n,i} + \hat m_i \mu_{m,i})(\hat n_j \mu_{n,j} + \hat m_j \mu_{m,j}) 
+ \left\{\hat a^\dagger_i \hat a_j  (\mu_{19}^2 w_i^* w_j  p^2 - \tfrac{1}{8} \mu_{35}^2 (s_i^* s_j + v_i^* v_j) q^2) + h.c.\right\} \nonumber \\
&&  \quad \quad \quad \quad \quad \quad + \left\{\hat b^\dagger_i \hat b_j  (\mu_{1,11}^2 W_i^* W_j  p^2 - \tfrac{1}{8} \mu_{83}^2 (S_i^* S_j + V_i^* V_j) q^2) + h.c.\right\}\Big] \nonumber \\
&& - \tfrac{3}{4} \sin^2 \theta \Big[e^{-2 i \phi}  \big\{ - \tfrac{1}{2} \mu_{35}^2 (\hat a^\dagger_i \hat a_j v^*_i s_j  + \hat a_i \hat a^\dagger_j v^*_j s_i) q^2 
 - \tfrac{1}{2} \mu_{83}^2 (\hat b^\dagger_i \hat b_j V^*_i S_j   + \hat b_i \hat b^\dagger_j V^*_j S_i) q^2\big\}
+ h.c.\Big],
\ea
\end{widetext}
where the second expression is obtained by making a further projection onto $\ket{\d} = \ket{0}$, $\ket{\u} = \hat a^\dagger \ket{0}$, and $\ket{\U} = \hat b^\dagger \ket{0}$. Here $\hat n_i = \hat a^\dagger_i \hat a_i$, $\hat m_i = \hat b^\dagger_i \hat b_i$,  $\mu_{n,i} = \mu_6 (|v_i|^2 - |s_i|^2) +\mu_{14} (|b_i|^2 - |a_i|^2)$, and $\mu_{m,i} = \mu_7 (|S_i|^2 - |V_i|^2) + \mu_{15} (|A_i|^2 - |B_i|^2)$.

To reduce the Hamiltonian governing $\hat a$ and $\hat b$ to two copies of hardcore bosons obtained from polar molecules in near-zero electric field \cite{yao13c}, we define $s_i q = r \tilde{x}_i$, $v_i q= r \tilde{y}_i$, $w_i p= r \tilde{z}_i |\mu_{35}/(2 \mu_{19})|$, $S_i q = |\mu_{35}/\mu_{83}|  r \tilde{X}_i$, $V_i q = |\mu_{35}/\mu_{83}|  r \tilde{Y}_i$, $W_i p = r \tilde{Z}_i |\mu_{35}/(2 \mu_{1,11})|$. We keep $|\tilde s_i|^2 + |\tilde v_i|^2 + |\tilde w_i|^2 = 1$ and $|\tilde S_i|^2 + |\tilde V_i|^2 + |\tilde W_i|^2 = 1$. $r$ is a positive real number. $c$ and $C$ are used to keep the states normalized. To mimic polar molecules, which are electric dipoles and thus have no induced dipole moments at small electric fields, we would like to choose $a_i$, $b_i$, $A_i$, and $B_i$ such that $\mu_{n,i} = \mu_{m,i} = 0$.   To do this, we reduce $r$ from $1$ and below until we find $p$ such that there is enough population in $a_i$, $b_i$, $A_i$, and $B_i$ levels to cancel dipole moments $\mu_{n,i}$ and $\mu_{m,i}$ for $\hat a$ and $\hat b$, respectively, for both sublattices. 
For example, for the parameters used to produce Fig.\ \ref{fig:c2}, $r = 0.265$ and $p = 0.953$.

The resulting Hamiltonian is: 
\begin{widetext}
\ba
\frac{4}{r^2 \mu_{35}^2} \hat H_{ij} & = & (1 \!-\! 3 \cos^2 \theta)  \Big[ \hat a^\dagger_i \hat a_j  (\tilde w_i^* \tilde w_j  - \tfrac{1}{2}  (\tilde s_i^* \tilde s_j + \tilde v_i^* \tilde v_j))  + \hat b^\dagger_i \hat b_j  (\tilde W_i^* \tilde W_j - \tfrac{1}{2} (\tilde S_i^* \tilde S_j + \tilde V_i^* \tilde V_j)) \Big] \nonumber \\
&& + \tfrac{3}{2} \sin^2 \theta \cos(2 \phi) \Big[  \hat a^\dagger_i \hat a_j (\tilde v^*_i \tilde s_j +  \tilde s_i^*  \tilde v_j) + \hat b^\dagger_i \hat b_j (\tilde V^*_i \tilde S_j +  \tilde S_i^*  \tilde V_j) \Big] \nonumber \\
&& - \tfrac{3}{2} \sin^2 \theta \sin(2 \phi) i \Big[  \hat a^\dagger_i \hat a_j (\tilde v^*_i \tilde s_j -  \tilde s_i^*  \tilde v_j) + \hat b^\dagger_i \hat b_j (\tilde V^*_i \tilde S_j -  \tilde S_i^*  \tilde V_j) \Big] + \textrm{h.c.}.
\ea
\end{widetext}
As desired, it has no density-density interactions, and both $\hat a$ and $\hat b$ obey the hopping Hamiltonian derived from polar molecules \cite{yao13c}.

In Fig.\ \ref{fig:c2}, we use $\Theta_0 = 0.68$ and $\Phi_0 = 5.83$ as the angles determining the direction of the DC electric field. The remaining parameters on the two sublattices (which we denote by  $1$ and $2$ in the subscript) are parametrized as $\tilde s_{1/2} = \tilde S_{1/2} = \sin(\tilde \alpha_{1/2}) \sin(\tilde \theta_{1/2})$, $\tilde v_{1/2} = \tilde V_{1/2} = \sin(\tilde \alpha_{1/2}) \cos(\tilde \theta_{1/2}) e^{i \tilde \phi_{1/2}}$, $\tilde w_{1/2} = \tilde W_{1/2} = \cos(\tilde \alpha_{1/2}) e^{i\tilde  \gamma_{1/2}}$ with $\{\tilde \theta_1, \tilde \theta_2, \tilde \phi_1,\tilde  \phi_2, \tilde \alpha_1, \tilde \alpha_2, \tilde \gamma_1,\tilde  \gamma_2\} =  \{0.53,  0.97, 1.36, 3.49,  2.84, 2.03, 4.26,  3.84\}$. As in the $C=1$ example, an additional minus sign is imposed on $w_{1/2}$ and $W_{1/2}$ on every other row. It is important to repeat that, while $\hat a$ and $\hat b$ here obey the same Hamiltonian, one should be shifted relative to the other by one lattice site in the $X$ direction. The parameters used in Fig.\ \ref{fig:c2diagram} are the same as those in Fig.\ \ref{fig:c2}, except $\Theta_0$ and $\Phi_0$ are varied.


%

\end{document}